# Multi Criteria Decision Making Approach for Selecting Effort Estimation Model

Sumeet Kaur Sehra
Assistant Professor
Guru Nanak Dev Engg. College, Ludhiana

Yadwinder Singh Brar
Professor
Guru Nanak Dev Engg.College, Ludhiana

Navdeep Kaur
Professor
Chandigarh Engineering College, Landran, Mohali

## ABSTRACT
Effort Estimation has always been a challenging task for the Project managers. Many researchers have tried to help them by creating different types of models. This has been already proved that none is successful for all types of projects and every type of environment. Analytic Hierarchy Process (AHP) has been identified as the tool that would help in Multi Criteria Decision Making. Researchers have identified that AHP can be used for the comparison of effort estimation of different models and techniques. But the problem with traditional AHP is its inability to deal with the imprecision and subjectivity in the pairwise comparison process. The motive of this paper is to propose Fuzzy Analytic Hierarchy Process, which can be used to rectify the subjectivity and imprecision of AHP and can be used for selecting the type of Model best suited for estimating the effort for a given problem type or environment. Instead of single crisp value, Fuzzy AHP uses a range of values to incorporate decision maker's uncertainty. From this range, decision maker can select the value that reflects his confidence and also he can specify his attitude like optimistic, pessimistic or moderate. In this work, the comparison of AHP and Fuzzy AHP is concluded using a case study of selection of effort estimation model.

## 1. INTRODUCTION
Software Development Effort estimation is related to prophecy of hours of work and number of workers needed to develop a project. The effort invested in a software project is probably one of the most important variables in the process of project management and is measured in Person-Months. Effort Estimates can be used to predict cost and duration estimates of the project. Effort estimates may be used as input to project plans, iteration plans, budgets, and investment analyses, pricing processes and bidding rounds.  Reliability in Effort Estimation is the biggest challenge for project Managers,

Jorgensen [1] analyzed that 60-80% of the software projects will overrun their estimated time and cost by 30%. Some other authors [2, 3, 4] also concluded that there is great need of good estimation techniques. There are many Software Effort Estimation models namely Algorithmic, Expert judgment based and Non Algorithmic.

Expert judgment relies purely on the experience of one or more experts. Algorithmic cost estimation involves the application of a cost model, i.e. one or more mathematical formulas which, typically, have been derived through statistical data analysis. Machine Learning techniques embody some of the facets of the human mind that allow us to solve hugely complex problems at speeds which outperform even the fastest computers [5]. Machine Learning techniques have been used successfully in solving many difficult problems such as speech recognition from text [6], adaptive control [7]; and mark-up estimation in the construction industry [8]. Recently Machine Learning approaches have been proposed as an alternative way of predicting software effort [9]. All of the three approaches have known advantages and disadvantages. Many effort estimators support one ideology, for example, either expert judgment or model-based estimates. Some conduct more than one estimate, using one to support the other. Others combine estimates from multiple approaches. So it can be interpreted that no single technique is best for all [10]. Meli and Menzies [11, 12] have suggested such an integrated approach by combining expert judgment with models.

Analytic Hierarchy Process has been identified as a tool that could help in Multi criteria Decision Making Problem such as effort estimation. But the potential drawback with the AHP [13] method is "Rank Reversal". Because judgments in AHP are relative by nature, changing the set of alternatives may change the decision scores of all the alternatives. As if a new, very poor alternative is added to a completed model, the alternatives with top scores sometimes reverse their relative ranking [14].

The motive of this paper is to propose Fuzzy Analytic Hierarchy Process Multicriteria Decision Making which compares the characteristics of expert judgment, algorithmic model and non-algorithmic models systems for type of project or environment of software Development Company.  The concept of Expert Judgment, Model Based approaches, fuzzy hierarchical analytic approach for Multicriteria decision-making (MCDM) is also discussed in the subsequent section. Then an illustrative example is presented, applying the MCDM methods, after which discussion related to effectiveness of MCDM methods is presented. In this work, the decision makers have been selected from the small companies in and around Chandigarh, Punjab. |they are asked to do the necessary required comparison between the different components of the models as would be discussed in the proceeding sections. Finally, the conclusions are presented and future directions for the modeling of the software effort estimation discussed.

## 2. EFFORT ESTIMATION MODELS
### 2.1 Expert Judgment
The expert judgment method involves consultation with one or more local experts who have knowledge about the development environment or application domain to estimate the effort required to complete a software project based on the environment. The estimate based on expert judgment where the process of generating an estimate is not explicit or recoverable, and is based largely on the intuition of experts [4]. Jorgensen supports expert predictions by claiming that formal models should not replace expert judgment, but should instead support expert prediction [1]. Most of the software industry is much more willing to accept, understand, and properly use judgment-based estimation methods [15].

The experts called upon for estimation in expert judgment input are expert software engineers, or senior project managers [16] but this is not mandatory that they are going to be good estimators also [17]. Expert judgment fails in many cases either giving underoptimsitic or overoptimistic estimates. A study is conducted by [18] in the specific areas in which expert judgment both succeeds and fails, and what commonalities underpinned the disciplines that consistently displayed successful and un-successful expert judgment.

A methodology is proposed by [19] to improve the expert judgment. The methodology is based upon the findings of Jorgensen [20] and Bolger and Wright [18]. The methodology is divided into three phases: record, learn and estimate, this methodology suggests that with increase in the size of a work item, the application potential and the importance of the estimation increase. According to Jarabek [19], larger work items should be estimated with a more top-down estimation strategy with more information about the estimation process being recorded. Smaller work items should be estimated with a bottom-up estimation strategy but less information should be recorded about each work item in order to reduce unnecessary overhead.

## 2.2 Model Based

Model-based methods uses algorithm to take past data as input and make estimations about new projects. Most Estimation models are based on the size measure such as Line of Code (LOC) and Function Point (FP). The accuracy of size estimation directly influences the accuracy of effort estimation.

Software effort prediction models fall into two main categories: algorithmic and non-algorithmic. The most popular algorithmic estimation models include Boehm's COCOMO [21], Putnam's SLIM [22] and Albrecht's Function Point. The algorithmic models have difficulty in modeling the inherent complex relationships between the contributing factors, are unable to handle categorical data as well as lack of reasoning capabilities [23]. The limitations of algorithmic models led to the exploration of the non-algorithmic techniques which are soft computing based. These include artificial neural network, evolutionary computation, fuzzy logic models, case-based reasoning, and combinational models [24]. The advantage of using a model based approach is that the historical data available can be used to evaluate the effectiveness of the model [25].

A Model is proposed by [26] to use Particle Swarm Optimization to tune the simple COCOMO model parameters such that better software effort estimation is achieved. The performance of the developed model is evaluated using NASA software projects data set [27].Out of 18 projects in data set, 13 projects are used as training set and the 5 projects in the testing (i.e. validating) set. The evaluation Criteria used is MMRE. MMRE value obtained in this model is quite less than the already existing models. The research is also going on to use ant colony optimization for effort estimation of the software products.

## 3. FUZZY MULTICRITERIA DECISION MAKING

The MCDM method deals with the process of making decisions in the presence of multiple criteria or objectives. A decision maker (DM) is required to choose among quantifiable or non-quantifiable and multiple criteria. The DM's evaluations on qualitative criteria are often subjective and imprecise. The objectives are usually conflicting and therefore the solution is highly dependent on the preferences of the DM [28]. Besides, it is very difficult to develop a selection criterion that can precisely describe the preference of one alternative over another. The evaluation data of subject alternatives suitability for various subjective criteria, and the weights of the criteria are generally expressed in linguistic terms.

As there is strong need of combining the different models [12], the approach that could be used out of many available approaches is Analytic hierarchy process [11, 29, 30]. But inability of AHP is to deal with the imprecision and subjectivity in the pairwise comparison process, which would improve in Fuzzy AHP [31]. Instead of single crisp value, Fuzzy AHP will use a range of values to incorporate decision maker's uncertainty. From this range, decision maker can select the value that reflects his confidence and also he can specify his attitude like optimistic, pessimistic or moderate [4]. Optimistic attitude is represented by the highest value of range, moderate attitude is represented by the middle value of the range and pessimistic attitude is represented by the lowest value of the range.

## 3.1 Analytic Hierarchy Process

Analytic Hierarchy Process (AHP) proposed by Satty, is an approach for decision making that involves structuring multiple choice criteria into a hierarchy, assessing the relative importance of these criteria, comparing alternatives for each criterion, and determining an overall ranking of the alternatives on the basis of cost, benefits and risk [30]. In AHP, DM is required to provide judgments about the relative importance of each criterion and then he has to specify a preference for each decision alternative on each criterion. The output of the AHP is prioritized ranking indicating the overall preference for each of the decision alternatives eventually help the decision maker to select the best approach.. The AHP consists of the following four steps:

1. Decision hierarchy is constructed by breaking down the decision problem into a hierarchy of inter-related and interdependent elements.
2. Pairwise comparisons are made on the decision elements.
3. Estimation of the weights of elements is done by using Eigen method.
4. The weights of the decision elements are aggregated to provide a set of ratings for the decision alternatives.

Starting with, the decision problem is broken into a hierarchy of interrelated & interdependent decision elements. Fig.1 exemplifies such a hierarchy. First nod of the hierarchy is the most general objective of the problem such as the objective of making the best decision or selecting the best alternative. The number of Criteria levels or sub levels depends on the complexity of the problem and on the degree of detail. Iterating from the more general to the more specific level, problem is split into sub-modules that will become sub-hierarchies. Navigating through the hierarchy from top to bottom, the AHP structure comprises goals, criteria and alternative ratings Each branch is then further divided into an appropriate level of detail. At the end, the iteration process transforms the unstructured problem into a manageable problem organized both vertically and horizontally under the form of a hierarchy of weighted criteria. By increasing the number of criteria, the importance of each criterion is thus diluted, which is compensated by assigning a weight to each criterion [30].

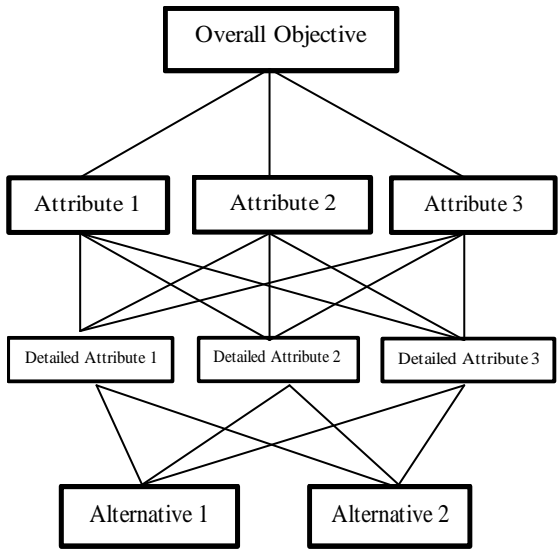

**Fig. 1 Structure of Analytic Hierarchy Process**

Then a relative weight is assigned to each criterion, based on its importance within the node to which it belongs. The sum of all the criteria belonging to a common direct parent criterion in the same hierarchy level must equal 100% or 1. A global priority is computed that quantifies the relative importance of a criterion within the overall decision made [30].

The alternatives are scored and compared to each other. Using AHP, a relative score for each alternative is assigned to each node within the hierarchy, and then to the branch of the node belongs to, and so on, up to the top of the hierarchy, where an overall score is computed. The degree of importance is measured on nine-point scale. The scale from 1 to 9 is given in the Table 1. The importance ratios for each pair of alternatives, a matrix of pairwise comparison ratios is then obtained. The criterion being compared may have different importance as compared to each other. So a pairwise comparison matrix is considered for the one criterion. Values of this matrix are pairwise or mutual importance ratios between the criteria, the values are decided on the basis that how important it is in evaluating the final goal The comparison is done in the form of pair of values and placed in matrix A.

$$A = \begin{bmatrix} 1 & a12 & \cdots & a1n \\ 1/a12 & 1 & \cdots & a2n \\ \vdots & \vdots & & \vdots \\ 1/a1n & 1/a2n & \vdots & 1 \end{bmatrix}$$

In the above matrix A the values of the elements tells that how much dominating the $i^{th}$ element on $j^{th}$ element.
1. $a_{ij} = 1/a_{ji}$, for $a_{ij} \neq 0$
2. $a_{ij} = 1$, for $i = j$ and $i, j = 1, 2, \ldots, n$.

A is a reciprocal matrix. The DM has the option of expressing preferences between the two as equally preferred, weakly preferred, strongly preferred or absolutely preferred, which would be translated into pair-wise weights of 1,3,5,7 and 9 respectively, with 2, 4, 6, and 8 as intermediate values [30]. Depending upon the judgment of the DM, if all the comparisons are perfect then matrix A is referred as Consistency Matrix.

The Principle of eigenvalues of matrix A is used to give the consistency of the judgments made. Every matrix has a set of

**Table 1:- Saaty's scale for pairwise comparison [30]**

| Saaty's scale | The relative importance of the two sub-elements |
|---|---|
| 1 | Equally important |
| 3 | Moderately important with one over another |
| 5 | Strongly important |
| 7 | Very strongly important |
| 9 | Extremely important |
| 2,4,6,8 | Intermediate values |

eigenvalues and for every eigenvalue there is a corresponding eigenvector, which is a vector of relative weights and defined as
$$w = [w_1 \ w_2 \ldots w_n]^T \quad (1)$$

The AHP method use two techniques to determine the final weights; one is Lambda Max($\lambda_{max}$) technique and second is geometric mean. While calculating the weights using $\lambda_{max}$ method, every matrix developed has a set of eigenvalues and every eigenvalue there is an eigenvector. [30] defines a vector of weights as the normalized eigenvector corresponding to the largest eigenvalue $\lambda_{max}$), then the following formula shows how weights are calculated.

$$Aw \cong nw \Rightarrow (A - \lambda_{max} I)w = 0 \Rightarrow \lambda_{max} = \frac{1}{n}\sum_{i=1}^{n}\frac{(Aw)_i}{w_i} \quad (2)$$

Where A is comparison matrix, w is weight matrix and $\lambda_{max}$ is eigenvalues.

The second method used for calculating the weights is Geometric mean method [31], in this method following formula is used to calculate the weights.

$$r_i = (\prod_{j=1}^{n} a_{ij})^{1/n} \Rightarrow w_i = r_i / \sum_{i=1}^{n} r_i \quad (3)$$

where n is the number of alternatives and $a_{ij}$ is the comparison ratios of matrix A. In practical situation it not possible for DM to make 100% correct judgment when n is large. For evaluation of the performance of consistency of the decision maker, Consistency Index is given as

$$C.I. = (\lambda_{max} - n)/(n-1) \Rightarrow w = (w_1, w_2, \ldots, w_n) \quad (4)$$

where n is the number of alternatives or criteria and $\lambda_{max}$ is the maximum value of the eigenvalues.

## 3.2 Fuzzy Analytic Hierarchy Process

There is a problem with AHP that in some situations, Decision maker wants to use the uncertainty while performing the comparisons of the alternatives [31]. For taking uncertainties into consider ration fuzzy numbers are used instead of crisp numbers. Instead of single crisp value, Fuzzy

## 3.2.1 Fuzzy Numbers

Lotfi Zadeh [32] proposed, elegant approach to deal with vagueness called fuzzy set theory. In his approach an element can belong to a set to a degree k ($0 \leq k \leq 1$), in contrast to classical set theory where an element must definitely belong or not to a set [32]. The degree or extent to which the elements are the members of the interval known as membership function is given as (5). The triangular membership function is given in Fig.2.

$$\mu_A(x) = \begin{cases} 0 \; for\, x \leq l \\ \dfrac{x-l}{m-l} \; for\, l \leq x \leq m \\ \dfrac{u-x}{u-m} \; for\, m \leq x \leq u \\ 0 \; for\, x \geq n \end{cases} \quad (5)$$

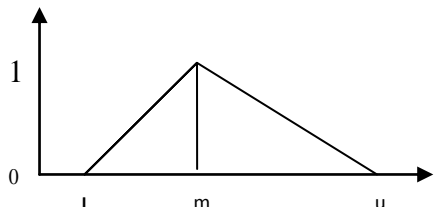

**Fig. 2 Triangular Membership Function**

## 3.2.2 Calculation of Fuzzy Weights from FAHP Comparison Matrices

As discussed above in the AHP, [31] was the first one to discuss the method for calculating the Fuzzy weights. The idea of using fuzzy triangular number is to give the DM an opportunity that he can decide in better way if there is little uncertainty in deciding the dominance of one alternative over the other. Now Triangular Fuzzy Number (TFN) is given as

$a_{ij} = (l_{ij}, m_{ij}, u_{ij})$ (6)

Here l is the lower limit value, m is the most promising value and u is the upper limit value. So fuzzy comparison matrix differs with Saaty's scaling in which membership scales are used instead of the 1-9 scales, as in Table 2, as the values of the elements.

Using the values of scales and fuzziness given in Table 2, a fuzzy comparison matrix A= $(a_{ij})_{nxn}$ where i, j = 1,2,3…… n is constructed. But the level of fuzziness or the TFN value of an element in matrix A decided by the DM based on the Saaty's values presented in Table 1 & fuzziness ranges of Table 2. After obtaining the fuzzy matrix, now priority weights are calculated. To obtain the estimate of the weight values for each criterion and for each alternative with reference to a given criteria. So firstly the synthetic extent values are obtained which is given as under

$$S_i = \sum_{j=1}^{m} N_{ci}^j \otimes \left[ \sum_{i=1}^{n} \sum_{j=1}^{m} N_{ci}^j \right]^{-1} \quad (7)$$

Where $N_{ci}^j$, j = 1, 2, 3….n are TFN values and $\otimes$ is fuzzy multiplication operation. The degree of possibility of $N_1 \geq N_2$ is defined as,

$$V(N_1 \geq N_2) = \sup_{x \geq y}[\min(\mu_{N_1}(x), \mu_{N_2}(y))] \quad (8)$$

When a pair $(x, y)$ exists such that $x \geq y$ and $\mu_{N_1}(x) = \mu_{N_2}(y) = 1$, then $V(N_1 \geq N_2) = 1$. Since the numbers $N_1$ and $N_2$ are convex fuzzy numbers so,

$V(N_1 \geq N_2) = 1$ if $n_{11} \geq n_{21}$ (9)

$V(N_2 \geq N_1) = hgt(N_1 \cap N_2) = \mu_{N_1}(d)$ (10)

Where $d$ is ordinate of the highest intersection point $D$ between $\mu_{N_1}$ and $\mu_{N_2}$. When $N_1$ and $N_2$ are fuzzy numbers then ordinate of D is computed as

$$V(N_2 \geq N_1) = hgt(N_1 \cap N_2) = \dfrac{l_1 - u_2}{(m_2 - l_2) - (m_1 - l_1)} \quad (11)$$

For the comparison of $N_1$ and $N_2$, both the values of $V(N_1 \geq N_2)$ and $V(N_2 \geq N_1)$ are required. The degree possibility for a convex fuzzy number to be greater than k convex fuzzy numbers $N_i (i = 1,2,...,k)$ is defined by [36]

$V(N \geq N_1, N_2,...,N_k) = V[(N \geq N_1),...,(N \geq N_k)] = \min V(N \geq N_i)$ (12)

If $m(A_i) = \min V(S_i \geq S_k)$, for $k = 1,2,...,n$; $k \neq i$, then the weight vector $W_p$ is given by, $W_A = (m(A_1), m(A_2),...,m(A_n))^T$ where $A_i(i=1,2,...,n)$ are $n$ elements. In order to normalize the weight vector $W_A$ the following formula is used.

$$W_A = \dfrac{W^T}{\left(\sum W^T\right)} \quad (13)$$

## 4. CALCULATING WEIGHTS FOR DIFFERENT EFFORT ESTIMATION MODELS

First stage is the identification of the necessary criteria, which would be used to evaluate different models are compared to each other. The criteria identified are given follows:

a. Reliability ($C_1$):- It tells about the behavior of the models corresponding to different situations. It differs from hardware reliability in that it reflects the design perfection, rather than manufacturing perfection.

b. Mean Magnitude of Relative Error (MMRE) ($C_2$) :- It measures the relative error absolute value of mean of [(actual - estimate)/actual]

c. Pred(x/100) ($C_3$):-: It gives the percentage of projects for which estimate is within x% of the actual

d. Uncertainty ($C_4$):- It tells in case of small errors, how the estimates changes.

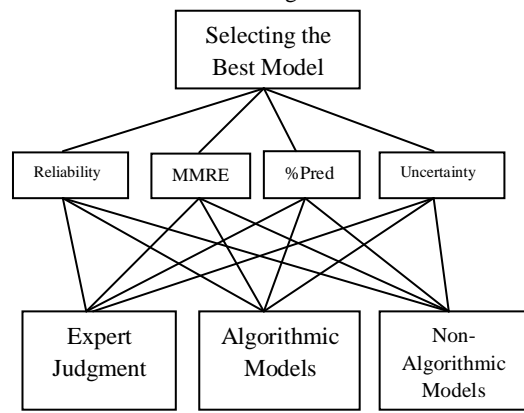

**Fig 3. The hierarchy of effort estimation Models**

After selecting the criteria, Decision Alternatives (DA) are also considered, which are, Expert Judgment, COCOMO, Fuzzy Neural Network based effort estimation models (hereafter $A_1$, $A_2$, $A_3$). The decisions Maker are taken from the software development companies in and around city Chandigarh, Punjab, India. After providing the necessary information regarding criteria and decision alternatives the Decision Makers (DMs) are asked to make preferences between the pair criteria and then pair of alternatives over the different criteria based on the previous knowledge of projects and evaluation of the work of the researchers. The decision Makers are first asked to fill a questionnaire about the criterion and alternative and then after it matrices table 3 and table 4 are created, using the linguistic variables as discussed Above in table no. 1, which used to make pairwise comparison. As discussed by the [34] there should not be any biasing or pressure on the DM to compare the criterion, which he doesn't want to compare. It is also difficult to give the some of the preferences as crisp values so for such uncertainty the fuzzy comparison matrix is developed. In this fuzzy comparison matrix, the $Z_{ij}$, as discussed in previous section, is the fuzzy number which is obtained after comparing ith DA with jth DA. So table 5 and 6 are created using Triangular Fuzzy numbers.

**Table 2:- Scale for fuzzy pairwise comparison [33]**

| Linguistic scale for importance | Fuzzy numbers for fuzzy AHP | Membership function | Domain | Triangular fuzzy scale (l, m, u) |
|---|---|---|---|---|
| Just equal | | | | (1.0, 1.0, 1.0) |
| Equal importance | 1 | $\mu M(x) = (3 - x)/(3 - 1)$ | $1 \leq x \leq 3$ | (1.0, 1.0, 3.0) |
| Weak importance of one over another | 3 | $\mu M(x) = (x - 1)/(3 - 1)$ $\mu M(x) = (5 - x)/(5 - 3)$ | $1 \leq x \leq 3$ $3 \leq x \leq 5$ | (1.0, 3.0, 5.0) |
| Essential or strong importance | 5 | $\mu M(x) = (x - 3)/(5 - 3)$ $\mu M(x) = (7 - x)/(7 - 5)$ | $3 \leq x \leq 5$ $5 \leq x \leq 7$ | (3.0, 5.0, 7.0) |
| Very strong Importance | 7 | $\mu M(x) = (x - 5)/(7 - 5)$ $\mu M x = (9 - x)/(9 - 7)$ | $5 \leq x \leq 7$ $7 \leq x \leq 9$ | (5.0, 7.0, 9.0) |
| Extremely preferred | 9 | $\mu M x = (x - 7)/(9 - 7)$ | $7 \leq x \leq 9$ | (7.0, 9.0, 9.0) |
| If factor i has one of the above numbers assigned to it when compared to factor j, then j has the reciprocal value when compared with I | | | | Reciprocals of above $(1/u1, 1/m1, 1/l1)$ |

**Table 3:- Pairwise comparisons for given criterion**

|    | C1 | C2  | C3  | C4 |
|----|----|-----|-----|----|
| C1 | 1  | 1/5 | 1/3 | 1  |
| C2 | 5  | 1   | 1   | 7  |
| C3 | 3  | 1   | 1   | 7  |
| C4 | 1  | 1/7 | 1/7 | 1  |

**Table 4:- Pairwise comparisons of alternatives with respect to criteria**

| C1 | A1 | A2  | A3  |
|----|----|-----|-----|
| A1 | 1  | 1/5 | 1/5 |
| A2 | 5  | 1   | 3   |
| A3 | 5  | 1/3 | 1   |

(a)

| C2 | A1 | A2 | A3  |
|----|----|----|-----|
| A1 | 1  | 1  | 1   |
| A2 | 1  | 1  | 1/3 |
| A3 | 1  | 3  | 1   |

(b)

| C3 | A1 | A2 | A3  |
|----|----|----|-----|
| A1 | 1  | 1  | 1   |
| A2 | 1  | 1  | 1/3 |
| A3 | 1  | 3  | 1   |

(c)

| C4 | A1  | A2  | A3  |
|----|-----|-----|-----|
| A1 | 1   | 5   | 5   |
| A2 | 1/5 | 1   | 1/7 |
| A3 | 1/3 | 1/7 | 1   |

(d)

**Table 5:- Fuzzy matrix of the criteria with respect to Goal**

|    | C1        | C2                | C3              | C4     |
|----|-----------|-------------------|-----------------|--------|
| C1 | [1,1,1]   | [0.14, 0.2, 0.33] | [0.2, 0.33, 1]  | [1, 1, 1] |
| C2 | [3, 5, 7] | [1,1,1]           | [1,1,1]         | [5,7,9] |
| C3 | [1, 3, 5] | [1,1,1]           | [1,1,1]         | [5,7,9] |
| C4 | [1, 1, 1] | [0.11, 0.143, 0.2]| [0.11, 0.143, 0.2] | [1,1,1] |

**Table 6:- Fuzzy Pairwise comparisons of alternatives with respect to criteria**

| C1 | A1 | A2 | A3 |
|----|----|----|----|
| A1 | [1, 1, 1] | [0.14, 0.2, 0.33] | [0.14, 0.2, 0.33] |
| A2 | [3, 5, 7] | [1,1,1] | [1, 3, 5] |
| A3 | [3, 5, 7] | [0.2, 0.33, 1] | [1,1,1] |

| C2 | A1 | A2 | A3 |
|----|----|----|----|
| A1 | [1, 1, 1] | [1, 1, 1] | [1, 1, 1] |
| A2 | [1, 1, 1] | [1, 1, 1] | [0.2, 0.33, 1] |
| A3 | [1, 1, 1] | [1, 3, 5] | [1, 1, 1] |

| C3 | A1 | A2 | A3 |
|----|----|----|----|
| A1 | [1, 1, 1] | [1, 1, 1] | [1, 1, 1] |
| A2 | [1, 1, 1] | [1, 1, 1] | [0.2, 0.33, 1] |
| A3 | [1, 1, 1] | [1, 3, 5] | [1, 1, 1] |

| C4 | A1 | A2 | A3 |
|----|----|----|----|
| A1 | [1, 1, 1] | [3, 5, 7] | [3, 5, 7] |
| A2 | [0.14, 0.2, 0.32] | [1,1,1] | [5, 7,9 ] |
| A3 | [0.14, 0.2, 0.33] | [0.11, 0.2, 0.14] | [1, 1, 1] |

The first stage of the weight evaluation process is the aggregation of $l_{ij}$, $m_{ij}$ and $u_{ij}$ values present in the pairwise comparison matrix for the judgments between criteria, calculated as the row sums and column sums as in table 7.

**Table 7:- Sum of rows and columns based on different criteria**

|    | Rows Sum | Column Sum |
|----|----------|------------|
| C1 | (2.343,2.53,3.53) | (6,10,14) |
| C2 | (10,14,18) | (2.253,2.343,2.53) |
| C3 | ( 8,12.16) | (2.31,2.473,3.2) |
| C4 | (2.22,2.286,2.4) | (12,16,20) |
| Sum of Column Sums | | (22.563,30.816,39.73) |

The associated $S_i$ synthetic extent is calculated for given criterion is given as

$S_1 = \left(\frac{2.343}{39.73}, \frac{2.53}{30.816}, \frac{3.23}{22.563}\right)$ = (0.0589, 0.0821, 0.147)

$S_2 = \left(\frac{10}{39.73}, \frac{14}{30.816}, \frac{18}{22.563}\right)$ = (0.25, 0.45, 0.8)

$S_3 = \left(\frac{8}{39.73}, \frac{12}{30.816}, \frac{16}{22.563}\right)$ = (0.20, 0.39, 0.71)

$S_4 = \left(\frac{2.22}{39.73}, \frac{2.286}{30.816}, \frac{2.4}{22.563}\right)$ = (0.056, 0.074, 0.11)

Using Equations (11) and (12), raw weights are calculated using synthetic extent calculated above given in table 8.

**Table 8: Raw Weights based on different criteria**

| Synthetic Index | | | | | Raw weights |
|---|---|---|---|---|---|
| | $S_1$ | $S_2$ | $S_3$ | $S_4$ | |
| V($S_1 \geq S_2, S_3, S_4$) | - | 0.512 | 1.8 | 1 | 0.512 |
| V($S_2 \geq S_1, S_3, S_4$) | 1 | - | 1 | 1 | 1 |
| V($S_3 \geq S_1, S_2, S_4$) | 1 | 1.11 | - | 1 | 1 |
| V($S_4 \geq S_1, S_2, S_3$) | 1.1 | 2.02 | 1.93 | - | 1.1 |

After normalization of the weights, the weights of $C_1$, $C_2$, $C_3$ and $C_4$ are as W = (0.14, 0.28, 0.28, 0.30). Similar mathematical procedure is used for calculating the weights for given criteria. The outcome of procedure is given as in table 9.

**Table 9:- Set of weights for all fuzzy matrices**

|  | Weights of Decision Alternatives | | | Weights for Criteria |
| --- | --- | --- | --- | --- |
|  | $A_1$ | $A_2$ | $A_3$ |  |
| $C_1$ | 0.44 | 0.28 | 0.28 | 0.14 |
| $C_2$ | 0.286 | 0.43 | 0.286 | 0.28 |
| $C_3$ | 0.286 | 0.43 | 0.286 | 0.28 |
| $C_4$ | 0.26 | 0. | 0.48 | 0.30 |
| Final Results | 0.2918 | 0.3580 | 0.3434 |  |

The results in table 9 reveal that the Decision Maker prefers MMRE and Pred criterion and also prefers the COCOMO model for the estimation of software development effort.

## 5. COMPARISON OF AHP AND FUZZY AHP

Analytic hierarchy process is the best way to conclude based on number of decisions and whereas the fuzzy AHP is the synthetic extension of AHP, when the fuzziness of the decisions has to be considered. In classical AHP these numerical values are exact numbers whereas in fuzzy AHP method they are intervals between two numbers with most likely value. As the nature of the human being, linguistic values can change from person to person. In these circumstances, taking the fuzziness into account will provide less risky decisions.

Based on the recommendations of the decision maker the tables 3 & 4 have been created using classical AHP the results of the calculations are given in table 10.

**Table 10:- Set of weights using Classical AHP**

| Model Used | Normalized Weights using Classical AHP | Normalized Weights using Fuzzy AHP |
| --- | --- | --- |
| Algorithmic Model | 0.243598 | 0.3580 |
| Expert Judgment | .320006 | 0.2918 |
| Non-Algorithmic Model | 0.436396 | 0.3434. |

Table 10 results are clearly showing that Non-algorithmic Model, which has weight factor of 0.4363, and has clear dominance over other models according to the classical AHP approach whereas Algorithmic Model having weight factor of 0.3580 has dominance over the others which are based on fuzzy AHP..

So there are two different results using two approaches. The results are basically showing that there was a margin of fuzziness that affected the overall result. Here the point that should not be missed, classical and fuzzy method are not the competitors with each other at same conditions. The important point is that if the information is certain then classical method should be preferred; if the information is not certain then fuzzy AHP method should be preferred. In this case study better results are in case of FAHP as compared to classical AHP as per the recommendation of the Decision Makers.

## 6. CONCLUSION

Researchers on software effort estimation have concluded that no practice for finding the effort and cost estimation of the software project is best. But different approaches have different success rates in different environments. The aim of this work is to examine the application of the Fuzzy Analytic Hierarchy Process (FAHP) method of multi-criteria decision-making for selecting the best model based on the company environment and type of the project. For doing the same DM has to select the best mode. The important consequences of the choice outcome may confer a level of uncertainty on the decision maker, in the form of doubt, procrastination etc. This is one reason for the utilization of FAHP, with its allowance for imprecision in the judgments made.

Future work will be to work with the decisions, when DM does not want to make any comparison between any two criterion/alternatives and can leave that comparison matrix entry empty. Efforts can be made to implement other Multi Criteria Decision Making approaches, using Different Fuzzy Numbers and optimization of the weights of the MCDM.

## 7. REFRENCES